\documentclass[aps,prb,twocolumn,%
reprint,showpacs,showkeys]{revtex4-1}

\usepackage{graphicx}
\usepackage{dcolumn}
\usepackage{bm}
\usepackage{textcomp,comment}
\usepackage[usenames]{color}

\begin{document}

\preprint{Phys. Rev. B}

\title{Adsorbate-Mediated Growth of Rare-Earth Oxides on Silicon}

\author{Bj\"orn Kaemena}

\author{Jan Ingo Flege} \email{flege@ifp.uni-bremen.de}%
 
\author{Jens Falta}

\affiliation{%
   Institute of Solid State Physics, University of Bremen,
   Otto-Hahn-Allee 1, 28359 Bremen, Germany}%

\date{\today}

\begin{abstract}
  Ultrathin cerium oxide films have been deposited on chlorine,
  gallium, and silver passivated Si(111) by reactive molecular beam
  epitaxy in a comparative study. The crystallinity of these films has
  been characterized by x-ray standing waves while the oxidation state
  of the rare-earth oxide (REOx) films and the chemical interface
  composition have been revealed by hard x-ray photoelectron
  spectroscopy.  The use of Cl as passivating agent results in the
  epitaxial growth of highly crystalline REOx films with the RE metal
  in the 3+ oxidation state while effectively suppressing silicate and
  silicon oxide formation at the interface.  In contrast, Ga and Ag
  preadsorption yield films of inferior quality, in the case of Ag of
  even lower crystallinity than without passivation.  Further
  investigations show that Cl-passivation also results in ultrathin
  lanthana films of superior quality, which facilitate the growth of
  well-ordered ceria on lanthana REOx multilayers.
\end{abstract}





\maketitle

\section{Introduction}

Due to their intriguing physical and chemical properties, which mainly
arise from the unfilled $4f$ electron shell, rare-earth oxides (REOx)
have been studied intensely in order to understand the nature of these
extraordinary properties, with the aim of exploiting them in various
kinds of technological applications.  Besides their traditional
utilization in fields like catalysis,\cite{Trovarelli_01} some of the
REOx have attracted a lot of interest in microelectronics due to their
high static dielectric constants $k$, which in the case of, e.\,g.,
CeO$_2$ amounts to $k>26$,\cite{Nishikawa_APL_2002,
  Nishikawa_JJAP_2002} their large band gap ($\sim6$\,eV for CeO$_2$
(Refs.~\onlinecite{Koelling_SolidStateCommun_1983,
  Wuilloud_PhysRevLett_1984})), and their predicted thermodynamic
stability in contact with silicon,\cite{Schlom_J_Mater_Res_1996}
making them promising candidates for novel "high-$k$" dielectrics.

To maintain the progression of integration densities in
microelectronics, there is a growing desire for ultrathin,
well-ordered, epitaxial REOx layers with a sharp interface to the
underlying silicon substrate as a replacement for the SiO$_2$ gate
oxide. These novel gate oxide layers would allow for further
downscaling of field-effect transistors (FET) by simultaneously
increasing device performance and lowering power
consumption.\cite{Huff_Gilmer_2004} However, the interface between the
silicon substrate and the high-$k$ gate oxide requires precise
engineering since the interface trap densities and the carrier
scattering need to be minimized to allow for reliable,
high-performance devices.\cite{Wilk_JAP_2001} Therefore, almost perfect
epitaxial interfaces between the silicon substrate and the high-$k$
material are required.  Yet, achieving a well-defined epitaxial
interface has so far been considerably impeded by RE-promoted silicon
oxidation, resulting in amorphous silicon oxide and RE silicate
formation.\cite{Hillebrecht_PRB_1986, Henle_PRB_1990,
  Chikyow_ApplPhysLett_1994, Preisler_JVSTB_2001, Narayanan_APL_2002}

To overcome these challenges given by the high reactivity of the
REOx-silicon interface different approaches have been employed,
ranging from influencing the growth kinetics by varying the REOx
growth rate and substrate temperature to introducing surface active
agents as, e.\,g., hydrogen.\cite{Yoshimoto_JpnJApplPhys_1995,
  Hirschauer_ASS_1999, Nishikawa_APL_2002} Since hydrogen limits the
growth temperature to 450\,\textcelsius\ due to its relatively low
desorption temperature,\cite{Gupta_PRB_1988} other adsorbates, which
allow for higher growth temperatures, would in principle be more
favorable because at given interface stability higher growth
temperatures directly translate into an increased crystallinity of the
deposited REOx film.  In this respect, chlorine presents a suitable
alternative, which is commonly used in semiconductor
processing.\cite{Yu_SurfSciRep_1994}

Recently,\cite{Flege_Kaemena_PRB_2011} we have suggested the use of
chlorine passivation for the growth of well-ordered Ce$_2$O$_3$
adlayers on Si(111) by reactive molecular beam epitaxy (MBE) in
ultra-high vacuum (UHV).  Here, we widen the scope and test the
general concept of REOx growth following substrate pre-passivation,
which we will call adsorbate-mediated growth, in an extensive study
employing several commonly used surface active agents. First, we
compare the crystalline quality, interface composition and oxidation
state of ultrathin cerium oxide films deposited on chlorine, silver,
and gallium passivated silicon(111) by employing X-ray standing waves
(XSW) and hard X-ray photoelectron spectroscopy (HAXPES). For Cl
passivation, which yields the cerium oxide films of the highest
quality, we also perform XSW-HAXPES with O$1s$ photoelectrons in a
chemically sensitive manner, thereby shedding light on the atomic
interface structure of silicon oxide and silicate species.

In the second part of the manuscript, we test the transferability of
the concept of Cl passivation to other REOx by investigating its
influence on the growth of lanthana films on bare and chlorine
passivated Si(111), again by utilizing XSW and HAXPES to quantify the
crystallinity of ultrathin lanthana films and the interface
composition.  XSW also allows monitoring of the Cl binding sites
during the growth process, again by employing XSW and XPS.  Together
with the cerium oxide results, this analysis unambiguously confirms
the beneficial effect of Cl preadsorption with respect to REOx
crystallinity and interface sharpness.

Finally, in the last part we demonstrate that the adsorbate-mediated
growth concept enables the realization of well-defined ultrathin REOx
multilayers, i.\,e., a well-ordered, ultrathin cerium oxide layer on
top of a highly crystalline, ultrathin lanthana film that was grown on
Cl-passivated Si(111). By using XSW we are able to selectively probe
and quantify the crystallinity of each REOx layer, while HAXPES serves
to elucidate the chemistry at the lanthana-silicon interface and to
determine the oxidation state of each REOx layer.

\section{Experimental}

All experiments were performed at the Hamburg Synchrotron Radiation
Laboratory (HASYLAB), which is located at the Deutsches
Elektronensynchrotron (DESY) in Hamburg, Germany.  At the undulator
beamline BW1, the samples were prepared and characterized \emph{in
  situ} under UHV conditions employing XSW, XPS, and low-energy
electron diffraction (LEED). The XSW experiments were performed in a
nondispersive setup employing a water-cooled Si(111) double-crystal
monochromator with an asymmetrically cut second crystal for enhanced
phase contast.  As inelastic secondary signals, photoelectrons were
recorded at photon energies of 2.6\,keV and 3.35\,keV while x-ray
fluorescence data was collected at primary energies of 5.9\,keV. The
HAXPES measurements were conducted at photon energies of 3.35\,keV and
2.6\,keV.

Sample preparation started from polished, RCA-cleaned Si(111)
crystals, which were introduced into the UHV chamber and degased at a
temperature of 630\textcelsius\ for at least 12\,h. The Si(111)
crystals were then annealed to 880\textcelsius\ to remove the
protective silicon oxide from the RCA treatment and to achieve a
(7$\times$7) reconstruction, which was verified by LEED.  In the case
of Si(111) surface passivation chlorine, silver, and gallium were used
to saturate silicon dangling bonds prior to REOx growth.  In reference
experiments, REOx films were directly deposited onto the (7$\times$7)
reconstructed Si(111) surface. In the case of Cl passivation we
followed our recipe published earlier.  \cite{Flege_Kaemena_PRB_2011}
Passivation with Ag was achieved by saturation exposure of Ag from a
knudsen cell at 500\textcelsius\ substrate temperature leading to a
$(\sqrt{3}\times\sqrt{3})R30$\textdegree\
reconstruction\cite{Wan_PhysRevB_1993} as witnessed by LEED (data not
shown).  Ga passivation was achieved by evaporating metallic Ga using
an electron-beam evaporator at 625\textcelsius\ substrate temperature,
yielding a surface that exhibited a LEED pattern indicating the
presence of a prevailing $(\sqrt{3}\times\sqrt{3})R30$\textdegree\
reconstruction\cite{Kawazu_PhysRevB_1988, Patel_JVacSciTechnolB_1989}
together with a faint contribution of an incommensurate
(6.3$\times$6.3) phase.\cite{Zegenhagen_PhysRevB_1988,
  Patel_JVacSciTechnolB_1989} The very existence of the latter domains
ensures that essentially all dangling bonds of the Si substrate are
saturated because the (6.3$\times$6.3) phase is known to only form
after the completion of the $(\sqrt{3}\times\sqrt{3})R30$\textdegree\
phase in this temperature range.\cite{Gangopadhyay_unpublished} The RE
oxides were then deposited at the same conditions by evaporating the
RE metal using an electron-beam evaporator in a preset oxygen ambient
of $5\times10^{-7}$\,mbar partial pressure and a substrate temperature
of 500\textcelsius. Typical growth rates were in the range of
2\,\AA/min as determined from complementary x-ray reflectometry (XRR)
measurements.

In an XSW experiment, the sample reflectivity and the intensity of an
element-specific inelastic signal, e.\,g., x-ray fluorescence or
photoelectrons, are recorded simultaneously while tuning the sample
through the ($hkl$) Bragg condition.  By fitting the reflectivity and
the respective yield of the inelastic secondary signal within the
framework of the dynamical theory of x-ray diffraction
\cite{vL_60,Batterman_RMP_1964}, the modulus $f_c$ (''coherent
fraction'') and the phase $\phi_c$ (''coherent position'') of the
($hkl$) Fourier component of the spatial distribution function of the
contributing atoms may be determined with high
precision.\cite{Zegenhagen_JXST_1990} In general, for photoelectrons
as inelastic secondary signal non-dipole effects \cite{Woodruff_99,
  Schreiber_SSL_2001} may occur depending on the selected core level
and the experimental conditions. But, due to our measurement
geometry\cite{Flege_NJP_2005} these non-dipole effects are
minimized. A comparison between Ce$L_\alpha$ fluorescence and Ce$3d$
photoelectrons of a crystalline, one monolayer thin cerium oxide film
(data not shown) indicates non-dipole contributions of less than 8\,\%
in the coherent fraction. In the following, whenever fluorescence and
photolectrons as inelastic secondary signals from rare-earth atoms are
compared in an effort to quantify film crystallinity, we will correct
for non-dipole effects in the coherent fraction by taking into account
an 8\,\% difference.

As we have shown in our previous
publication,\cite{Flege_Kaemena_PRB_2011} XSW is highly suitable for
studying the epitaxial quality of ultrathin REOx films on silicon. A
quantitative measure for the crystallinity of a REOx film of known
thickness is the coherent fraction.  After correction for thermal
vibrations, its value can directly be compared to the theoretical
value of the coherent fraction calculated for given crystal structure
and film thickness.

\section{Results and Discussion}

In the following sections, we present our experimental results for
rare-earth oxide growth on passivated Si(111) surfaces. We will first
discuss the influence off different passivating agents
(Sec.~\ref{sec:passivating_agents_ceria}) on the epitaxial film
quality of ultrathin ceria films by employing HAXPES as well as
XSW-HAXPES, which complements and further enhances our previous
study.\cite{Flege_Kaemena_PRB_2011} In Sec.~\ref{sec:lanthana_7x7_cl}
we then also demonstrate the improved crystallinity of ultrathin
lanthana films due to chlorine passivation.  Finally, again employing
a combination of chemically sensitive XSW and hard x-ray photoelectron
spectroscopy (HAXPES) we assess the crystallinity and interface
composition of ultrathin rare-earth multilayers up to a few nanometers
on Cl-passivated Si(111) in Sec.~\ref{sec:multilayer}.

\subsection{Ceria growth on passivated Si(111)}
\label{sec:passivating_agents_ceria}

\begin{figure}
\includegraphics[width=0.9\linewidth]{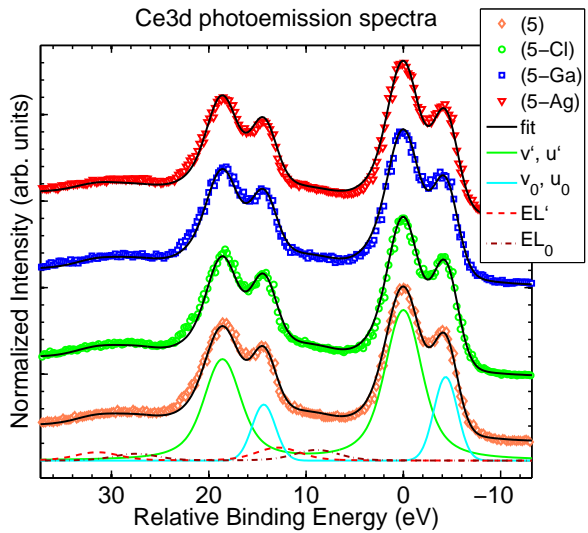}
\caption{\label{fig:XPS_surfactants_ce3d}(Color online) Ce$3d$
  photoemission spectra of ceria films with a film thickness of
  6\,\AA\ grown on non-passivated Si(111)-($7\times7$) and Cl, Ag and
  Ga passivated Si(111) at 500\,\textcelsius\ substrate temperature
  and an oxygen partial pressure of $5\times10^{-7}$\,mbar
  (preparation conditions: (5), (5-Cl), (5-Ag) and (5-Ga)).}
\end{figure}

We begin our discussion by comparing the oxidation state of cerium
oxide grown on Si(111) crystals passivated by Cl, Ag, and Ga as well
as cerium oxide grown on non-passivated Si(111). Since all oxide films
were grown at a substrate temperature of 500\textcelsius\ and an
oxygen background pressure of $5\times10^{-7}$\,mbar we introduce a
short hand notation for improved readability as follows: The
preparation conditions will be labeled by ''(5)'', ''(5-Cl)'',
''(5-Ag)'' and ''(5-Ga)'' identifying the oxygen partial pressure
(divided by $10^{-7}$\,mbar) and the passivating agent used in the
respective experiment. In Fig.~\ref{fig:XPS_surfactants_ce3d} we
present Ce$3d$ photoemission spectra of cerium oxide films for the
growth recipes (5), (5-Cl), (5-Ag) and (5-Ga) for a nominal film
thickness of 6\,\AA. Qualitatively, the four spectra show the same
peak structure and exhibit the typical shape that is attributed to a
single Ce$^{3+}$ (Ce$_2$O$_3$) oxidation
state.\cite{Mullins_SurfSci_1998} Quantitatively, all four spectra can
be fitted consistently with the established double peak structure
($u_0$,$u'$,$v_0$,$v'$) of Ce$_2$O$_3$ arising from the hybridized
final state.\cite{Kotani_AdvPhys_1988} The findings from the Ce$3d$
spectra reveal that the oxidation state of cerium oxide is Ce$^{3+}$
and that is does not depend on the applied passivating agent.

\begin{figure}
\includegraphics[width=0.99\linewidth]{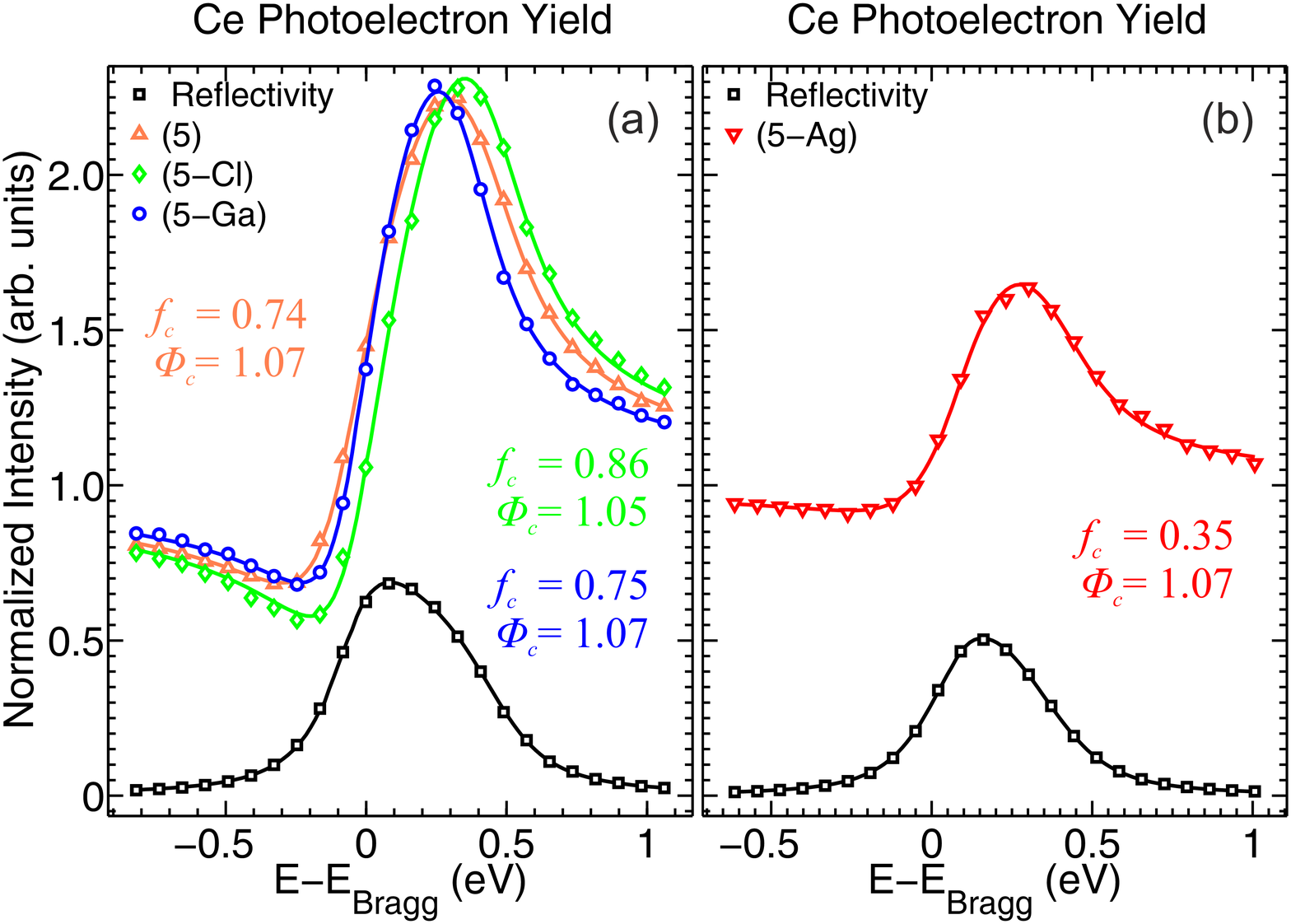}
\caption{\label{fig:xsw_ceria_passivation}(Color online) XSW data
  (data points) and theoretical fit according to the dynamical theory
  of x-ray diffraction (solid lines) using Ce$3d_{5/2}$ photoelectrons
  obtained in Si(111) Bragg reflection for 6\,\AA\ thick ceria films
  grown at 500\,\textcelsius\ substrate temperature and an oxygen
  partial pressure of $5\times10^{-7}$\,mbar on (a) Cl passivated
  Si(111) (5-Cl), Ga passivated Si(111) (5-Ga) and (b) Ag passivated
  Si(111).}
\end{figure}

In Fig.~\ref{fig:xsw_ceria_passivation}\,(a) and (b)%
the according XSW data are displayed for the 6\,\AA\ thick ceria films
(5), (5-Cl), (5-Ga) and (5-Ag) discussed above. The figures show the
reflectivity and the corresponding yields of the inelastic secondary
signals, which in this case are Ce$3d_{5/2}$ photoelectrons, acquired
at 3.35\,keV (a) and 2.6\,keV (b) incident photon energy,
respectively. The coherent position $\Phi_c$ of all four yields is the
same within the errorbar and on average amounts to
$\Phi_c=1.065$. Since the coherent position is independent of the
passivating agent applied prior to cerium oxide growth and all cerium
oxide films exhibit the same film thickness, the coherent fraction
$f_c$ is a direct measure of the crystalline order of the
film.\cite{Flege_Kaemena_PRB_2011} For the preparation conditions (5),
(5-Cl), (5-Ga), and (5-Ag) the coherent fractions amount to
$f_c\text{(5)}=0.74$, $f_c\text{(5-Cl)}=0.86$, $f_c\text{(5-Ga)}=0.75$
and $f_c\text{(5-Ag)}=0.35$, respectively. Therefore, we conclude that
chlorine passivation of Si(111) prior to cerium oxide growth leads to
the highest crystalline quality of ultrathin ceria films while
passivation with silver drastically impedes crystallinity and leads to
much worse ordering than found for cerium oxide films deposited
without passivation. Gallium pre-adsorption does not seem to have any
effect on the crystalline order of the ceria film since the difference
in the coherent fraction compared to the ceria film grown on
non-passivated Si(111)-($7\times7$) is within the error bar.

In quantitative XSW simulations as described
earlier,\cite{Flege_Kaemena_PRB_2011} we assumed a laterally fully
strained Ce$_2$O$_3$ film of two O$_{0.75}$-Ce-O$_{0.75}$ trilayer
(TL) thickness with a relative spacing of 3.32\,\AA\ as derived from
continuum elasticity theory based on elastic constants for CeO$_2$.
\cite{Nakajima_PRB_1994} This model predicts a coherent fraction of
$f_c\text{(sim.)}=0.925$, which only differs by $\Delta f_c=0.065$
from the experimental result of $f_c\text{(5-Cl)}=0.86$ for the
Cl-passivated cerium oxide film.  Within the error bar of our XSW
measurement, this deviation can be fully attributed to thermal
vibrations.  Even when taking into account non-dipole contributions in
the order of a few percent, we conclude that the cerium oxide film
grown on Cl passivated Si(111) is very highly ordered.

\begin{figure}
\includegraphics[width=1\linewidth]{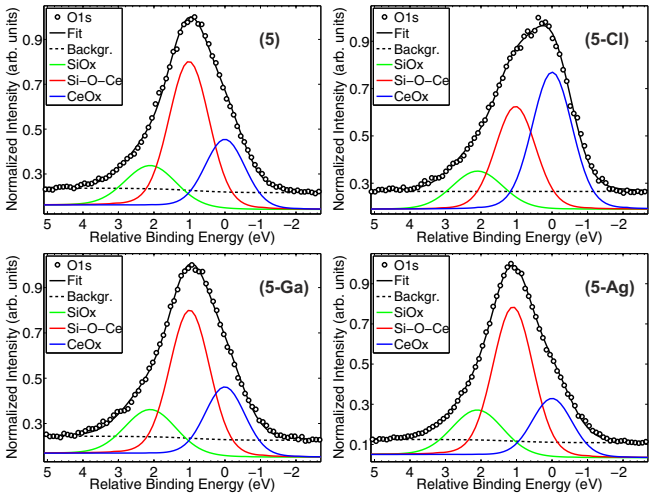}
\caption{\label{fig:xps_o1s_ceria_passivation}(Color online) O$1s$
  photoemission spectra of ceria films with a film thickness of
  6\,\AA\ grown on non-passivated Si(111) and Cl, Ag and Ga passivated
  Si(111) at 500\,\textcelsius\ substrate temperature and an oxygen
  partial pressure of $5\times10^{-7}$\,mbar (preparation conditions:
  (5), (5-Cl), (5-Ag), (5-Ga)). The spectra were recorded at photon
  energies of 3.35\,kev for ((5), (5-Cl) and (5-Ga)) and 2.6\,keV for
  (5-Ag) and have been deconvoluted intro three species: CeO$_x$,
  Si-O-Ce, SiO$_x$ using Voigt profiles.}
\end{figure}

After having discussed the influence of the passivating agent on the
epitaxial quality of the ceria films we now focus on the interface
composition by employing O$1s$ core level
spectroscopy. Fig.~\ref{fig:xps_o1s_ceria_passivation}%
shows the respective spectra for the samples discussed above (6\,\AA\
of cerium oxide grown at 500\,\textcelsius\ and $5\times10^{-7}$\,mbar
oxygen partial pressure on bare silicon Si(111)-($7\times7$) (5), Cl-
(5-Cl), Ga- (5-Ga), and Ag-passivated (5-Ag) Si(111)). The O$1s$
spectra have been deconvoluted into the following three species
\cite{Hirschauer_ASS_1999}: CeO$_x$, Si-O-Ce, SiO$_x$. The CeO$_x$
species belongs to the oxygen bound in the Ce$_2$O$_3$ lattice, the
Si-O-Ce species represents the oxygen that is bound to both silicon
and cerium atoms at the interface and the SiO$_x$ species stands for
the oxygen which oxidizes the silicon substrate. By just comparing the
relative intensities of the individual species it can easily be
deduced that only chlorine significantly decreases the interfacial
silicate and silicon oxide species. The other adsorbates gallium and
silver (O$1s$ spectra (5-Ga) and (5-Ag) in
Fig.~\ref{fig:xps_o1s_ceria_passivation}) do not show any significant
influence on the formation of amorphous interface species in
comparison with the cerium oxide growth on non-passivated
Si(111)-($7\times7$) (O$1s$ spectrum (5) in
Fig.~\ref{fig:xps_o1s_ceria_passivation}).

\begin{figure}
\includegraphics[width=0.99\linewidth]{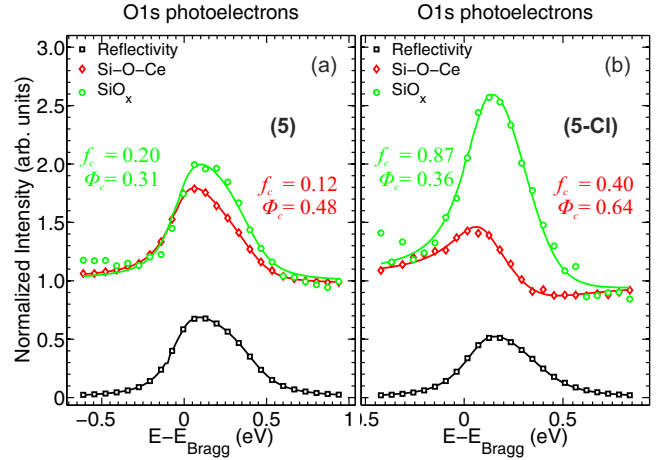}
\caption{\label{fig:XSW_O1s_ceria}(Color online) Chemical-sensitive
  XSW data (data points) and theoretical fit according to the
  dynamical theory of x-ray diffraction (solid lines) using O$1s$
  photoelctrons which belong to the Si-O-Ce and SiO$_x$ species
  obtained in Si(111) Bragg reflection for 6\,\AA\ thick ceria films
  grown at 500\,\textcelsius\ substrate temperature and an oxygen
  partial pressure of $5\times10^{-7}$\,mbar on (a) non-passivated
  Si(111)-($7\times7$) and (b) Cl passivated Si(111) (5-Cl).}
\end{figure}

Based on the previous O$1s$ HAXPES analysis we also performed
chemically-sensitive XSW in (111) Bragg reflection by monitoring the
respective yields of the chemically shifted O$1s$ components.  This
measurement provides quantitative insight into the spatial
distribution of the interfacial Si-O-Ce and SiO$_x$ species (confer
Fig.~\ref{fig:xps_o1s_ceria_passivation}) and facilitates a direct
comparison of their spatial coherence relative to the substrate
lattice.\footnote{Here, we note that analyzing the CeO$_x$ component
  in more detail does not provide direct information about the
  ordering of the oxygen ions within the film as coherent fractions
  close to $0$ are expected for highly ordered Ce$_2$O$_3$ bixbyite
  films independent of thickness.} The corresponding XSW data for the
preparation recipes (5) and (5-Cl) are depicted in
Fig.~\ref{fig:XSW_O1s_ceria}.%
As expected for amorphous phases, the coherent fractions
$f_c^{\text{Si-O-Ce}}\text{(5)}=0.12$ and
$f_c^{\text{SiO$_x$}}\text{(5)}=0.20$ obtained for the silicate and
silicon oxide species for the growth on bare Si(111) are very small
indicating a very poor ordering, probably due to several inequivalent
binding sites. But, a huge difference is observed for the coherent
fractions of the interfacial species in the case of Cl
passivation. While the analysis for silicate species Si-O-Ce yields a
considerably increased coherent fraction of
$f_c^{\text{Si-O-Ce}}\text{(5-Cl)}=0.40$ together with a coherent
position of $\Phi_c^{\text{Si-O-Ce}}\text{(5-Cl)}=0.64$, the silicon
oxide species at the interface seem to be very highly ordered with a
coherent fraction of $f_c^{\text{SiO}_x}\text{(5-Cl)}=0.87$ together
with a coherent position of
$\Phi_c^{\text{SiO}_x}\text{(5-Cl)}=0.36$. This is a direct result of
the low concentration of SiO$_x$ species
(Fig.~\ref{fig:xps_o1s_ceria_passivation}, (5-Cl)) as compared to the
Si-O-Ce and CeO$_x$ species, suggesting that only a small portion of
the interface exhibits silicon oxide species, which are incorporated
in an almost perfectly ordered fashion. Although the coherent fraction
of the Si-O-Ce species, which represent the majority of the
interfacial species, is more than tripled as compared to the recipe
(5), a suggestion of a structural model would be speculative at best
since at least a bimodal atomic distribution function would have to be
considered to satisfactorily explain its (111) Fourier
component. Nevertheless, our XSW results for both the SiO$_x$ and the
Si-O-Ce interface species offer a suitable benchmark within a combined
analysis using XSW and theoretical calculations, e.\,g., within the
framework of density functional theory,\cite{Flege_NJP_2005,
  Flege_PRB_2008} which could provide further insight into the complex
chemistry at the REOx-silicon interface.

Concluding the discussion of cerium oxide growth on Si(111), from our
XSW and HAXPES investigations we can unambiguously infer that chlorine
is the superior passivating agent for the realization of highly
ordered, epitaxial cerium oxide films on silicon accompanied by a very
low amount of ordered silicate and silicon oxide species at the
interface.  In the following section, we test the transferability of
the Cl passivation recipe for the growth of ultrathin lanthana films.

\subsection{Lanthana growth on Cl passivated Si(111)}
\label{sec:lanthana_7x7_cl}

\begin{figure}
\includegraphics[width=0.85\linewidth]{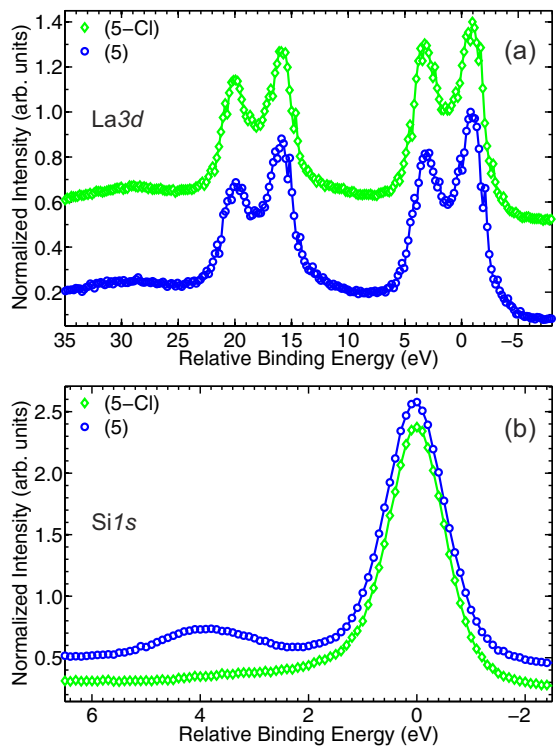}
\caption{\label{fig:xps_lanthana_si1s}(Color online) (a) La$3d$ and
  (b) Si$1s$ photoemission spectra of lanthana films with a film
  thickness of 1.8\,nm grown on Si(111)-($7\times7$) (open circles)
  and Cl/Si(111)-($1\times1$) (open diamonds).  Lanthana growth was
  performed at 500\,\textcelsius\ substrate temperature and an oxygen
  partial pressure of $5\times10^{-7}$\,mbar (preparation conditions:
  (5) and (5-Cl)). The spectra are vertically shifted, for clarity.}
\end{figure}

To study the influence of Si(111) surface passivation on other
rare-earth oxides besides ceria we prepared a 1.8\,nm thin lanthana
film on both bare Si(111)-$(7\times7)$ and Cl/Si(111)-($1\times1$).
In analogy to the previous experiments, these lanthana films were
grown at $5\times10^{-7}$\,mbar oxygen background pressure and a
substrate temperature of 500\,\textcelsius, again denoted by (5) and
(5-Cl). Fig.~\ref{fig:xps_lanthana_si1s} shows a representative La$3d$
photoemission spectrum and Si$1s$ spectra for preparation conditions
(5) and (5-Cl).%
As is visible from Fig.~\ref{fig:xps_lanthana_si1s}(a) the films were
grown as La$_2$O$_3$ since the spectrum exhibits the typical double
peak structure known for lanthana.\cite{Fuggle_PhysRevB_1983,
  Kotani_AdvPhys_1988} The effect of chlorine passivation is clearly
documented by the comparative Si$1s$ HAXPES data for the recipes (5)
and (5-Cl) (see Fig.~\ref{fig:xps_lanthana_si1s}(b)), which depending
on passivation exhibit an apparent intensity modulation in the range
of higher binding energies, i.\,e., $\sim 2-6$\,eV shifted relative to
the Si$1s$ main peak. In the (5-Cl) case, this part of the spectrum
only consists of inelastically scattered electron background and is
almost perfectly flat otherwise while the same region in the case of
(5) exhibits substantial integral intensity. This intensity can be
attributed to the presence of various silicon oxide species in
different oxidation states at the interface,\cite{Eickhoff_JESRP_2004}
e.\,g., silicon oxide and silicate.  Hence, from our spectroscopic
investigations we conclude that chlorine adsorption prior to lanthana
growth strongly reduces silicon oxidation at the interface.

\begin{figure}
\includegraphics[width=0.99\linewidth]{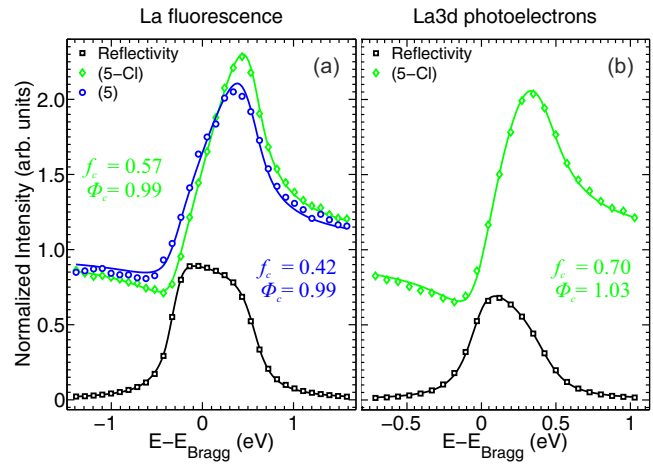}
\caption{\label{fig:xsw_lanthana}(Color online) XSW data (data points)
  and theoretical fit according to the dynamical theory of x-ray
  diffraction (solid lines) using (a) La$L_\alpha$ fluorescence and
  (b) La$3d_{5/2}$ photoelectrons obtained in Si(111) Bragg reflection
  for 1.8\,nm thick lanthana films grown at 500\,\textcelsius\
  substrate temperature and an oxygen partial pressure of
  $5\times10^{-7}$\,mbar on (a) non-passivated Si(111) (5) and (a),
  (b) Cl passivated Si(111) (5-Cl).}
\end{figure}

By employing XSW in (111) Bragg reflection
(Fig.~\ref{fig:xsw_lanthana}), we are able to assess and compare the
epitaxial quality of the lanthana films grown on passivated and
non-passivated silicon.
The La$L_\alpha$ fluorescence yields (see
Fig.~\ref{fig:xsw_lanthana}(a)) were recorded at an incident photon
energy of 5.9\,keV.  For the fluorescence data, the coherent position
$\Phi_c$ amounts to $\Phi_c=0.99$ for both preparation conditions. The
coherent fraction $f_c$ for the lanthana film grown after Cl
passivation amounts to $f_c\text{(5-Cl)}=0.57$ while the lanthana film
grown on Si(111)-($7\times7$) (5) only possesses a coherent fraction
of $f_c\text{(5)}=0.42$, yielding a significant difference of $\Delta
f_c=0.15$ in the coherent fractions. Therefore the XSW fluorescence
results reveal a higher crystallinity for the lanthana films grown on
chlorine passivated Si(111) as compared to lanthana deposited on bare
Si(111)-($7\times7$). This is in agreement with our XPS Si$1s$
investigation, which points toward a substantial concentration of
silicate and silicon oxide species at the interface for the growth
without passivation.

To perform an analogous quantitative XSW simulation for lanthana films
as in the case of cerium oxide,\cite{Flege_Kaemena_PRB_2011} the
crystallographic nature of the sesquioxide has been elucidated by
other experimental approaches.  Grazing incidence x-ray diffraction
studies of lanthana films grown on Cl-passivated silicon (data not
shown) reveal that La$_2$O$_3$, just like Ce$_2$O$_3$, crystallizes in
the so called C-type bixbyite structure and that the formation of the
well-known hexagonal phase is suppressed.  In quantitative XSW
simulations for a 1.8\,nm thick bixbyite lanthana film with a lattice
constant of $a=11.36$\,\AA\ (Ref.~\onlinecite{Adachi_ChemRev_1998}),
we assumed a laterally fully strained La$_2$O$_3$ film of five
O$_{0.75}$-La-O$_{0.75}$ trilayer thickness with a trilayer spacing of
3.45\,\AA\ as derived from continuum elasticity theory again based on
elastic constants for CeO$_2$ (Ref.~\onlinecite{Nakajima_PRB_1994}).
This model predicts a coherent fraction of
$f_c\text{(sim.)}=0.614$. If we assume an isotropic Debye-Waller
factor of $D^{(111)}=0.95$, then the simulated coherent fraction lies
within the error bar of the experimental value of
$f_c\text{(5-Cl)}=0.57$.  This result clearly indicates an almost
perfect crystalline order of the lanthana film grown on chlorine
passivated Si(111).

Further information on the atomic arrangement of the lanthanum atoms
perpendicular to the Si(111) diffraction planes in the lanthana film
(5-Cl) is gained by performing XSW measurements in (111) geometry
employing La$3d_{5/2}$ photoelectrons as secondary signal, which were
recorded at a photon energy of 3.35\,keV (see
Fig.~\ref{fig:xsw_lanthana}(b)). Because generally the escape depth of
photoelectrons in a certain material (in the order of \AA ngstr\"oms
to nanometers) is far less than the escape depth of photons (in the
order of several $\mu$m), the XSW photoelectron measurement is more
sensitive to the atomic ordering of the uppermost lanthana layers
while the x-ray fluorescence measurement equally probes the entire
film. The coherent fraction employing La$3d_{5/2}$ photoelectrons
amounts to $f_c^{\text{La}3d_{5/2}}\text{(5-Cl)}=0.70$ and is
substantially higher than the corresponding coherent fraction
$f_c^{\text{La}L_\alpha}\text{(5-Cl)}=0.57$.  Extrapolating from the
previous discussion for cerium oxide 3d photoelectrons, this
difference cannot be solely explained solely by non-dipole effects and
suggests a structural origin. This finding is a first indication for
the lanthana film exhibiting a higher crystalline order in the upper
layers than near the interface.  However, this difference is still
relatively small if one takes into account non-dipole effects on the
order of about 8\%.  Furthermore, the coherent fraction deduced from
the fluorescence data is close to the simulated value of
$f_c\text{(sim.)}=0.614$, the partial derivative of the crystalline
order of the lanthana film in the vertical direction is close to zero,
i.\,e., the variation in crystallinity along the vertical direction is
almost negligible. Concluding this discussion, we note that the
overall crystallinity of the lanthana film is very high, underlining
the beneficial effect of chlorine passivation on oxide crystallinity.

\begin{figure}
\includegraphics[width=0.99\linewidth]{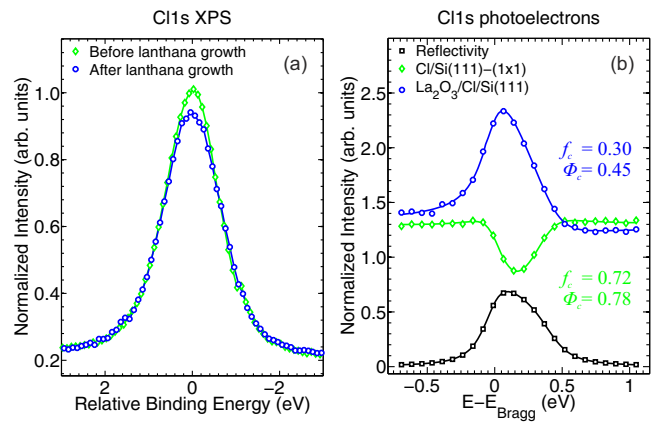}
\caption{\label{fig:XSW_XPS_Cl1s_lanthana}(Color online) (a) Cl$1s$
  photoemission spectra of chlorine-terminated Cl/Si(111)-($1\times1$)
  recorded at 3.35\,keV photon energy and after deposition of 1.8\,nm
  of lanthana grown at 500\,\textcelsius\ substrate temperature in an
  oxygen background pressure of $5\times10^{-7}$\,mbar. (b) XSW data
  (open symbols) obtained in Si(111) Bragg reflection using Cl$1s$
  photoelectrons and theoretical fit according to the dynamical theory
  of x-ray diffraction (solid lines) for the two preparation steps
  described in (a).}
\end{figure}

To study the behavior of chlorine in the lanthana growth process we
employed XPS and XSW on the Cl$1s$ core level before and after
lanthana deposition. Lanthana was deposited at 500\textcelsius\ and
on oxygen background pressure of
$5\times10^{-7}$\,mbar. Fig.~\ref{fig:XSW_XPS_Cl1s_lanthana}%
shows the XSW data and the normalized Cl$1s$ photoemission spectra
recorded at 3.35\,keV photon energy. The Cl$1s$ photoemission
intensity after lanthana growth is only slightly decreased. Since the
Cl$1s$ photoelectrons exhibit a kinetic energy of about 528\,eV, the
inelastic electron mean free path in La$_2$O$_3$ is approximately
1\,nm as derived from the TPP2M
formula.\cite{Tanuma_SurfInterfAnal_1994} Therefore, Cl$1s$
photoelectrons originating from chlorine atoms bound at the interface
of the 1.8\,nm thick lanthana film should be attenuated by a factor of
$~e^2$. From this analysis of the Cl$1s$ spectra we infer that the
chlorine atoms almost completely segregate to the surface of the
La$_2$O$_3$ film during growth. These findings are supported by the
Cl$1s$ XSW data, which are displayed in
Fig.~\ref{fig:XSW_XPS_Cl1s_lanthana}(b).  While the yield collected
before lanthana deposition essentially describes the expected Cl
on-top site identified in previous investigations,\cite{Flege_SS_2002,
  Flege_NJP_2005} we note a drastic change in the overall shape of the
inelastic secondary signal after lanthana growth, which quantitatively
results in variations from $f_c^{\text{Cl}1s}$ from $0.72$ to $0.30$
and $\Phi_c^{\text{Cl}1s}$ from $0.78$ to $0.45$ after oxide
growth. The strong decrease in the coherent fraction together with the
change in the coherent position prove a binding site change of the
chlorine atoms during growth, in total agreement with the XPS
results. Therefore we conclude that chlorine acts as surfactant and
that the interface passivation appears to be less stable. This finding
is in contrast to the role of chlorine during ceria growth at the same
preparation conditions, where chlorine was shown to mainly remain at
the interface, thereby acting as
interfactant.\cite{Flege_Kaemena_PRB_2011} Considering our earlier
results for praseodymia growth on Si(111) that indicated that Cl may
also act as a surfactant,\cite{Gevers_ApplPhysLett_2010} the growth
mechanism is an intricate interplay between the RE metal and the
adsorbate.

In conclusion the presented data for lanthana films grown at
preparation conditions (5-Cl) and (5) clearly show higher
crystallinity with strongly decreased silicate and silicon oxide
formation at the interface for the chlorine passivated
growth. Furthermore, a detailed analysis of the (5-Cl) lanthana film
proves that chlorine seggregates to the surface during growth.

\subsection{Rare-earth multilayers on Cl passivated Si(111)}
\label{sec:multilayer}

\begin{figure}
\includegraphics[width=0.99\linewidth]{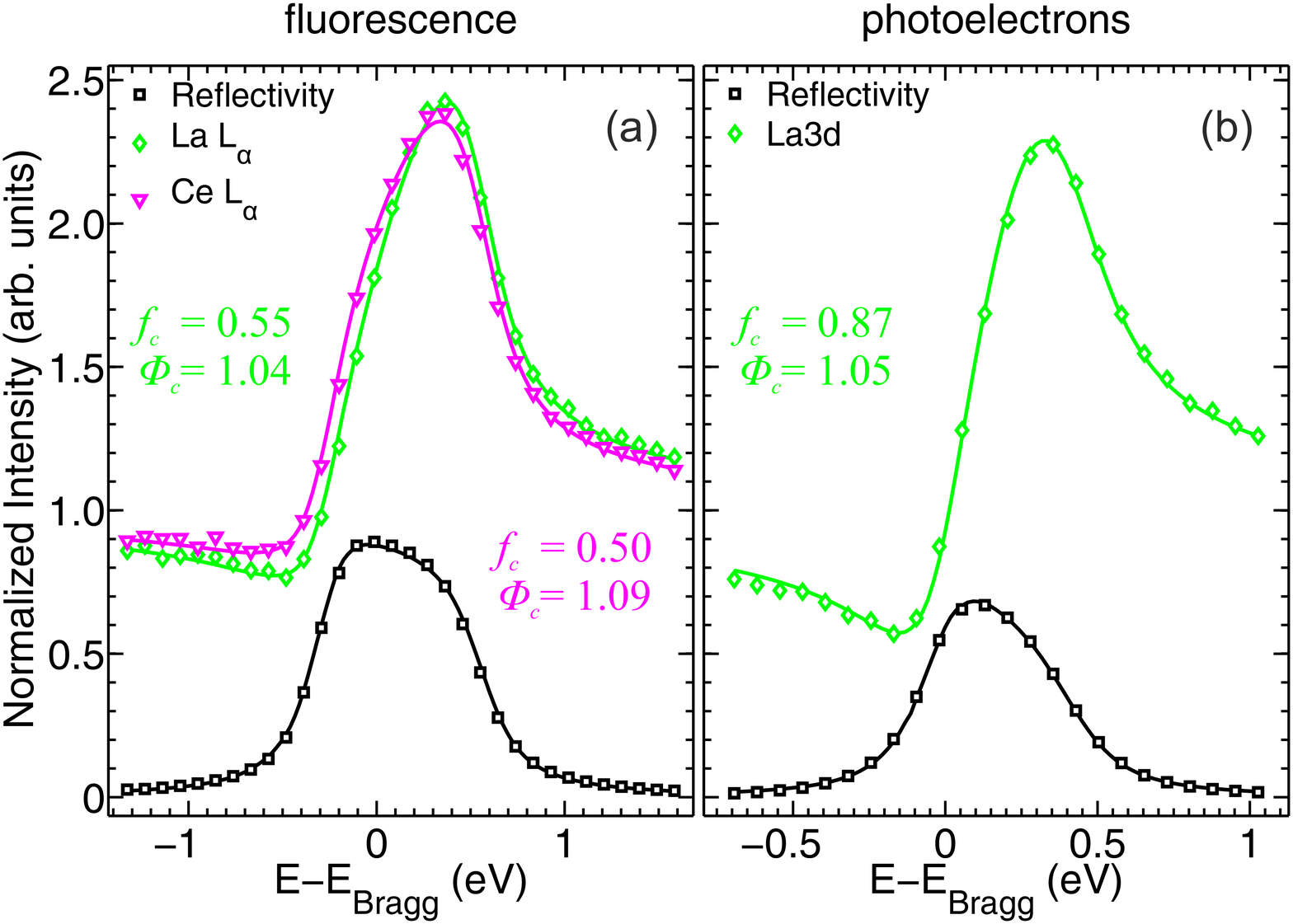}
\caption{\label{fig:xsw_multilayer}(Color online) XSW data (data
  points) and theoretical fit according to the dynamical theory of
  x-ray diffraction (solid lines) using (a) La$L_\alpha$ and
  Ce$L_\alpha$ fluorescence and (b) La$3d_{5/2}$ photoelectrons
  obtained in Si(111) Bragg reflection for a cerium oxide on lanthana
  rare-earth oxide multilayer film with individual layer thicknesses
  of 1.8\,nm for the lanthana film and 6\,\AA\ for the cerium oxide
  film grown at 500\,\textcelsius\ substrate temperature and an oxygen
  partial pressure of $5\times10^{-7}$\,mbar on Cl-passivated
  Si(111).}
\end{figure}

In this section we present first results for the growth of REOx
multilayers on chlorine-passivated Si(111) using the example of cerium
oxide growth on lanthana deposited on Cl/Si(111)-(1$\times$1). As
before in the single growth experiment
(Sec.~\ref{sec:passivating_agents_ceria}), the cerium oxide film was
grown at a substrate temperature of 500\textcelsius\ and an oxygen
background pressure of $5\times10^{-7}$\,mbar with a thickness of
6\,\AA. To facilitate a direct comparison with the previous
investigation (Sec.~\ref{sec:lanthana_7x7_cl}), the underlying
lanthana film was 1.8\,nm thick and also deposited at preparation
conditions (5-Cl). The corresponding XSW data are presented in
Fig.~\ref{fig:xsw_multilayer}. Again, fluorescence data acquired at
5.9\,keV and photoelectron data acquired at 3.35\,keV are shown in
Fig.~\ref{fig:xsw_multilayer}(a) and (b), respectively.%
The fit of the La$L_\alpha$ and Ce$L_\alpha$ yields results in
coherent fractions of $f_c^{\text{La}L_\alpha}=0.55$ and
$f_c^{\text{Ce}L_\alpha}=0.50$ with coherent positions of
$\Phi_c^{\text{La}L_\alpha}=1.04$ and
$\Phi_c^{\text{Ce}L_\alpha}=1.09$. The coherent fraction of the
Ce$L_\alpha$ signal is close to the coherent fraction of the
La$L_\alpha$ signal, which means that both REOx layers exhibit
comparable crystalline quality. But, the crystallinity of the cerium
oxide layer is somewhat inferior to the cerium oxide films grown
directly on Cl/Si(111)-(1$\times$1).\cite{Flege_Kaemena_PRB_2011}

When comparing the coherent fractions and the coherent positions of
the La$L_\alpha$ yields of the multilayer film and for the initial
lanthana layer on Cl/Si(111)-(1$\times$1), a change of
$\Delta\phi_c^{\text{La}L_\alpha}=0.05$ is observed in the coherent
position while the respective coherent fraction changes only by
$\Delta f_c^{\text{La}L_\alpha}=0.02$, meaning that only the coherent
position of the La$L_\alpha$ inelastic signal undergoes a
statistically significant, yet rather subtle change.  However, the
situation is completely different when employing La$3d_{5/2}$
photoelectrons as secondary signal
(Fig.~\ref{fig:xsw_multilayer}\,(b)), which exhibit a coherent
fraction of $f_c^{\text{La}3d_{5/2}}=0.87$ and a coherent position of
$\Phi_c^{\text{La}3d_{5/2}}=1.05$ after cerium oxide deposition. When
compared to the respective values before cerium oxide growth, these
findings reveal a substantial increase in the coherent fraction of
$\Delta f_c^{\text{La}3d_{5/2}}=0.17$ while the coherent position
stays the same within the error bar. Likewise, the difference in the
coherent fraction between the fluorescence data and the photoelectron
signal of the lanthana film in the rare-earth oxide multilayer is even
increased to $\Delta f_c=0.32$, which remains at more than $0.24$ even
if we again assume non-dipole effects in the order of $0.08$.  Hence,
these differences in the Fourier components for the individual
secondary signals can only be related to significant structural
changes within their respective sampling depths.  This indicates that
the top lanthana layers have become very highly ordered, as revealed
by the La$3d_{5/2}$ XSW measurement, as a side effect of the growth of
cerium oxide.  This ``healing'' of the topmost La$_2$O$_3$ layers has
to be counterbalanced by a respective loss of crystallinity in the
lowest layers near the interface, so that the overall epitaxial
quality of the lanthana film, which is monitored by the La$L_\alpha$
inelastic signal, may stay the same.

To get an even more detailed insight into this crystallinity gradient
of the lanthana film we performed quantitative XSW simulations taking
into account the finite escape depth of the La$3d$ photoelectrons with
a kinetic energy of 2.5\,keV in La$_2$O$_3$.  Applying the TPP2M
formula \cite{Tanuma_SurfInterfAnal_1994} yields a value of
$\lambda_{\text{La}_2\text{O}_3}(2.5\,\rm{keV})=3.95$\,nm. The
attenuation of the photoelectrons due to the cerium oxide film and the
associated change in probing depth can be neglected since the
corresponding attenuation factor
$e^{-d_{\rm{Ce}_2\rm{O}_3}/\lambda_{\rm{Ce}_2\rm{O}_3}}$ is the same
for all photoelectrons that are emitted from the lanthana
film. Therefore, the existence of the ceria film does not influence
the coherent fraction and coherent position of the La$3d$ XSW
results. The first result of the XSW simulations shows that the
calculated value for $\lambda_{\rm{La}_2\rm{O}_3}$ is too high to
explain the large difference of $\Delta f_c=0.24$ between the
fluorescence and the photolectron XSW data of a 1.8\,nm thick
La$_2$O$_3$ film. This is in accordance with spectroscopic
investigations of slightly thicker ceria films grown on Si(111), where
we find an inelastic electron mean free path that is about half of the
calculated value by the TPP2M formula \cite{Aschkan_inprep}. Hence, in
the following analysis we assume a value of
$\lambda_{\rm{La}_2\rm{O}_3}(2.5\,\rm{keV})/2$ and additionally take
into account the geometry of our experimental setup, where the
photoelectrons leave the sample in an off-perpendicular direction to
the surface leading to a further attenuation of photoelectrons. Our
XSW simulations reveal that the assumption of a laterally fully
strained La$_2$O$_3$ film with a trilayer spacing of 3.45\,\AA\ also
cannot explain the difference in the coherent fractions of the
fluorescence and the photoelectron data. In the following estimation
we assume a La$_2$O$_3$ film with 6 O-La-O trilayers with a smaller
trilayer spacing of 3.23\,\AA, which corresponds to a vertical lattice
constant expansion of 3\,\% compared to the vertical Si(111) lattice
periodicity.  A simple numerical model, which takes into account a
variation of crystalline order in $[hkl]$ direction, is then
implemented by associating each trilayer (TL) in the lanthana film
with a depth-dependent disorder parameter $C_i\in [0, 1]$, with $i$
representing the TL index as counted from the interface.  Within this
approach, a quite abrupt crystallinity gradient can be realized in the
following way: $C_{1}=0$, $C_{2}=0.05$, $C_3=0.5$, $C_{4}=1$,
$C_{5}=1$, and $C_{6}=1$.  Here, the extremal values 0 and 1 represent
total structural disorder and perfect bixbyite crystallinity,
respectively.  This model leads to coherent fractions of
$f_c^{\text{La}3d_{5/2},\rm{sim.}}=0.72$ and
$f_c^{\text{La}L\alpha,\rm{sim.}}=0.55$, i.\,e., a difference of
$\Delta f_c^{\rm{sim.}}=0.17$, with an according difference of the
coherent position of $\Delta\Phi_c^{\rm{sim.}}=0.01$, in agreement
with the experimental results for $\Delta\Phi_c$.

Although our simulations cannot completely reproduce a difference in
the coherent fractions of the photoelectron and fluorescence inelastic
secondary signal of $\Delta f_c=0.24$, it clearly reveals the presence
of perfectly ordered bixbyite trilayers in the upper part of the
La$_2$O$_3$ film in the rare-earth oxide multilayer structure and
essentially disordered layers close to the interface to the underlying
silicon. Although $\Delta f_c^{\text{sim}}$ does depend on the assumed
trilayer spacing as well as trilayer disorder, the result for the
crystallinity gradient always shows the following behavior: Few layers
at the interface are armorphous, while the upper layers are highly
ordered. The transition between total disorder and nearly perfect
crystallinity takes place in the regime of one or two trilayers and is
therefore very sharp.

\begin{figure}
\includegraphics[width=0.75\linewidth]{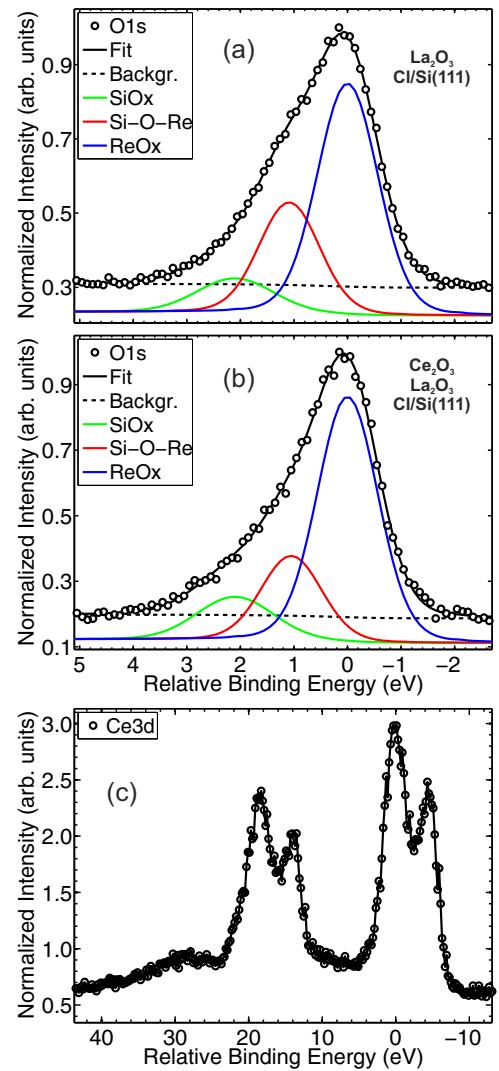}
\caption{\label{fig:xps_multilayer}(Color online) O$1s$ (a)
  photoemission spectra for a 1.8\,nm thick lanthana film grown at
  500\,\textcelsius\ substrate temperature and an oxygen partial
  pressure of $5\times10^{-7}$\,mbar on chlorine passivated
  Si(111). O$1s$ (b) and Ce$3d$ (c) photoemission spectra for a ceria
  on lanthana multilayer with a 6\,\AA\ thick ceria film deposited at
  500\,\textcelsius\ substrate temperature and an oxygen partial
  pressure of $5\times10^{-7}$\,mbar on-top of the lanthana layer of
  (a). The O$1s$ spectra (a) and (b) have been deconvoluted with Voigt
  profile fits including the following species: Re$_2$O$_3$, Si-O-Re,
  SiO$_x$.}
\end{figure}

From the quantitative XSW simulations of the La$_2$O$_3$ film we
conclude that upon the growth of cerium oxide at elevated temperatures
of 500\textcelsius\ and an oxygen background pressure of
$5\times10^{-7}$\,mbar the lanthana trilayers closer to the silicon
interface lose crystalline order while the upper trilayers in the film
become perfectly ordered.  This result is corroborated by the O$1s$
HAXPES data displayed in Fig.~\ref{fig:xps_multilayer}(a) and (b),
which were recorded at $3.35$\,keV incident photon energy for the
lanthana film grown on Cl/Si(111)-($1\times1$) and for the cerium
oxide on lanthana REOx multilayer, respectively.
The experimental spectra were fitted with three Voigt line shaped
oxygen species: an oxygen species in a RE$_2$O$_3$ chemical
environment (REO$_x$), a silicate species (Si-O-RE), and an silicon
oxide species (SiO$_x$). In the fitting procedure, the relative
binding energies as referred to the REO$_x$ species were assumed to be
shifted by 1.0\,eV (Si-O-RE) and 2.1\,eV (SiO$_x$), resulting in only
slight deviations from the values published for ceria grown on Si(111)
previously.\cite{Hirschauer_ASS_1999} The Lorentzian width was set to
0.15\,eV,\cite{Prince_JESRP_1999} while the Gaussian width was adapted
to fit the experimental resolution.  When normalizing the sum of all
species to 1, the integrated intensity of each species is $0.61$
(REO$_x$), $0.29$ (Si-O-RE), and $0.10$ (SiO$_x$) for the O$1s$
spectrum recorded prior to ceria growth
(Fig.~\ref{fig:xps_multilayer}\,(a)). After cerium oxide deposition
(Fig.~\ref{fig:xps_multilayer}\,(b)), the integrated intensities
amount to $0.65$ (REO$_x$), $0.22$ (Si-O-RE), and $0.14$
(SiO$_x$). Since the total film thickness is increased by a factor of
$~4/3$ from $1.8$\,nm to $2.4$\,nm due to the deposition of the cerium
oxide film, one would expect the following integrated intensities if
the amounts of Si-O-RE and SiO$_x$ species were not altered during
cerium oxide deposition and intensity attenuation due to a finite
escape depth of the O$1s$ photoelectrons are neglected: $0.68$
(REO$_x$), $0.24$ (Si-O-RE), $0.09$ (SiO$_x$). Although the
deconvolution of the O$1s$ spectrum results in an error of the
normalized intensity of approximately $\pm 0.01$, the deviations of
the actual intensities of the oxygen species after growth from the
expected intensities are significant and would be even larger if
taking into account the attenuation of the photoemission signal of the
interface species. This means that we do observe a ripening process at
the interface that is mainly accompanied by silicon oxide formation
during cerium oxide deposition due to elevated substrate temperatures
(500\,\textcelsius) and the exposure to oxygen at partial pressures of
$5\times10^{-7}$\,mbar. Hence, the increase of amorphous silicon oxide
species at the interface inferred from the HAXPES data is in very good
agreement with the decrease in crystalline order of the lanthana
layers at the interface, which is deduced from the analysis of the
coherent fraction of our XSW results. In addition, the increased
coherent position ($\Delta\Phi_c=0.05$) of the fluorescence XSW data
displayed in Fig.~\ref{fig:xsw_lanthana} and
Fig.~\ref{fig:xsw_multilayer} suggests that the lanthana film is
pushed slightly outwards as compared to the Si(111) diffraction
planes, which could be a side effect of the oxidation of the silicon
substrate.

After having discussed the epitaxial quality and the ripening of the
interface of the rare-earth multilayer we finally discuss the
spectroscopic investigation of the Ce$3d$ core level in
Fig.~\ref{fig:xps_multilayer}(c) of the ceria film, which is striking,
because it also reveals a Ce$^{3+}$ oxidation state. As it is known
ultrathin ceria films on silicon exhibit the Ce$^{3+}$ oxidation
state. The transition from the Ce$^{3+}$ to the Ce$^{4+}$ oxidation
state is supposed to take place with increasing film thickness while
the interfacial layers stay in the Ce$^{3+}$ oxidation
state.\cite{Hirschauer_TSF_1999} Since Ce catalyzes the formation of
silicon oxide and silicate,\cite{Hillebrecht_PRB_1986,
  Chikyow_ApplPhysLett_1994, Preisler_JVSTB_2001} it can be understood
why ceria on silicon exhibits the Ce$^{3+}$ oxidation state at the
interface and oxidizes with increasing film thickness. But here, we
present data of ultrathin epitaxial ceria in the Ce$^{3+}$ oxidation
state grown on La$_2$O$_3$, which should decouple the cerium oxide
film from the silicon support. Intriguingly, at the same preparation
conditions and for ceria islands up to 2\,nm, we find that ceria on
Ru(0001) can easily be oxidized.\cite{Kaemena_JPCC_2013} Hence, the
question arises why ceria on lanthana exhibits the Ce$^{3+}$ oxidation
state. While we cannot provide a definitive answer, a possible driving
force could be the very small lattice mismatch of C-type Ce$_2$O$_3$
($a=11.16$\,\AA) and C-type La$_2$O$_3$
($a=11.36$\,\AA),\cite{Adachi_ChemRev_1998} which would constrain the
ultrathin ceria film to pseudomorphic growth in the Ce$_2$O$_3$ phase.

To summarize this section, we have shown that ceria can be grown on
lanthana deposited on chlorine passivated Si(111) with comparable
epitaxial quality. Furthermore, it could be revealed that the lanthana
film becomes almost perfectly ordered in the upper layers during ceria
deposition, while the layers at the interface loose crystalline order
due to the ripening of silicon oxide species. Finally we demonstrated
that ultrathin ceria films also grow in the Ce$^{3+}$ oxidation state
on lanthana.

\section{Conclusion}

We have presented a detailed study on the adsorbate-mediated growth of
rare-earth oxides on Si(111). Based on synchrotron-based x-ray
photoemission and x-ray standing wave investigations we have
demonstrated that the specific type of chemical passivation of the
silicon substrate appears to be crucial in obtaining high-quality,
epitaxial rare-earth oxide films in molecular beam epitaxy.  Among the
passivating agents (Cl, Ag, Ga) tested, chlorine yielded the highest
degree of REOx film crystallinity combined with a well-ordered
oxide-semiconductor interface exhibiting only a very low concentration
of silicon oxide and silicate species.  Quite surprisingly, gallium
preadsorption yields the same degree of oxide crystallinity compared
to cerium oxide growth on the bare Si(111) surface, whereas
passivation with silver even drastically decreases the crystalline
order of the cerium oxide film.

Furthermore, the HAXPES and XSW results illustrate that the quality of
the cerium oxide films achieved is always tied to the structural and
chemical properties of the oxide-silicon interface.  Only in the case
of Cl, it is sharp and well-ordered with suppressed silicon oxide and
silicate species while in the other cases it was shown to comprise
considerable amounts of amorphous silicon oxide and silicate species.

The use of Cl passivation was also shown to be essential in obtaining
high-quality films of ultrathin lanthana on Si(111).  In this case,
the Cl seggregates to the surface of the lanthana film, rendering the
lanthana-silicon interface very susceptible to ripening during further
thermal treatment.  Hence, the role of chlorine as growth modifier
depends on the specific rare-earth oxide that is grown via molecular
beam epitaxy: Only in the case of cerium
oxide,\cite{Flege_Kaemena_PRB_2011} the Cl atoms mostly remain at the
interface, hence acting as \emph{interfactant}. However, for both
lanthana and praseodymia the Cl changes its adsorption site and takes
on the role as \emph{surfactant}.

The results for the cerium oxide on lanthana multilayer reveal that
both rare-earth oxide films exhibit comparable crystallinity with very
limited intermixing.  Furthermore, despite the coupling from the
substrate cerium oxide grows in the Ce$^{3+}$ oxidation
state. Detailed X-ray standing wave simulations of the rare-earth
oxide multilayer show that a ripening process occurs at the interface
between lanthana and silicon, which is documented by an increase in
amorphous silicon oxide species and which leads to a vertical
crystallinity gradient in the lanthana film.  However, cerium oxide
growth is also concomitant with an almost perfect ordering of the
upper lanthana bixbyite trilayers.

The superior crystallinity of rare-earth oxide films when grown on
Cl-passivated Si(111) is accompanied with a sharp and ordered
interface with suppressed silicate and silicon oxide species.  In
terms of application, this way of tuning the structural properties of
an epitaxial oxide-silicon interface might represent an important step
toward low interface trap densities and minimized carrier scattering,
eventually allowing for their integration as high-$k$ gate oxides.

\acknowledgments Portions of this research were carried out at the
light source DORIS III at DESY. DESY is a member of the Helmholtz
Association (HGF). We would like to thank D.\ Novikov for assistance in
using beamline BW1.


%

\end{document}